\documentclass[letterpaper]{article}
\usepackage{aaai}
\usepackage{times}
\usepackage{helvet}
\usepackage{courier}
\usepackage{todonotes}
\usepackage{lscape}

\usepackage{booktabs}
\usepackage{natbib}
\usepackage{hyperref}

\begin{document}
%
\title{An influencer-based approach to understanding radical right viral tweets}
\author{Laila Sprejer$^{1}$, Helen Margetts $^{1}$, Kleber Oliveira$^{2}$, David O'Sullivan$^{2}$, Bertie Vidgen$^{1\ddagger}$ \\
  {$^1$The Alan Turing Institute,
  $^2$MASCI, University of Limerick} \\
  {$^\ddagger$bvidgen@turing.ac.uk}\\
}

\maketitle
\begin{abstract}
Radical right influencers routinely use social media to spread highly divisive, disruptive and anti-democratic messages. 
Assessing and countering the challenge that such content poses is crucial for ensuring that online spaces remain open, safe and accessible.
Previous work has paid little attention to understanding factors associated with radical right content that goes viral.
We investigate this issue with a new dataset (R\textsc{ot}) which provides insight into the content, engagement and followership of a set of 35 radical right influencers. It includes over 50,000 original entries and over 40 million retweets, quotes, replies and mentions.
We use a multilevel model to measure engagement with tweets, which are nested in each influencer.
We show that it is crucial to account for the influencer-level structure, and find evidence of the importance of both influencer- and content-level factors, including the number of followers each influencer has, the type of content (original posts, quotes and replies), the length and toxicity of content, and whether influencers request retweets.
We make R\textsc{ot} available for other researchers to use.
\end{abstract}



\section{Introduction}
Incendiary, divisive and extremist online messages have the potential to cause real harm to society by disrupting civic discourse, exacerbating social tensions and spreading anti-democratic ideas. 
This content is particularly concerning when it is produced by ``influencers'' who not only have social power over their large, engaged followings but drive often-vitriolic opposition.
Prominent right-wing figures such as Nigel Farage, Paul Joseph Watson and Charlie Kirk have attracted substantial attention for their online posts, especially when they go ``viral'', reaching large audiences and attracting considerable engagement.
Prior research suggests that online spaces are turbulent and chaotic, which makes it difficult to explain why some content receives widespread engagement whilst other content attracts little attention \citep*{Margetts2015}. This in turn makes it difficult to plan and deliver strategic counter-communications and offer timely support to vulnerable communities. Platforms can also find it difficult to decide on appropriate interventions for hazardous content, risking public backlash.
However, little research to date has systematically considered which factors associated with radical right content facilitate the most engagement.
In particular, the role of individual content-creators has not been extensively studied.

We investigate which factors are associated with the highest levels of engagement for content produced by radical right wing influencers. We focus on a set of 35 influencers (including individuals, political parties, and other organisations) whose influence is directed towards or significant within the United Kingdom (UK). 
We address one primary research question: \textbf{Which factors are associated with the highest engagement rates for radical right content on Twitter?}
We show that both influencer- and content-level factors are important for explaining the engagement patterns of radical right Twitter content. We show that an influencer's total number of followers makes a substantial difference, and provide mixed evidence on whether follower composition makes a difference. We also demonstrate the importance of content-level variables, showing that type of content (original posts, quotes and replies), length and toxicity of content, inclusion of additional media, and retweet requests by influencers all make a difference.

We introduce the Radical Right On Twitter dataset (R\textsc{ot}). It covers the 186 days from 8 July 2020—9 January 2021, and contains (1) 90,675 tweets produced by 35 radical right influencers, including original content, quotes, replies and retweets, (2) 31,443,828 tweets engaged with that influencer content, and (3) 12,204,060 tweets which otherwise mention those influencers.
The dataset also lists each influencers' followers, collected each day, and the followers of every account which engaged with at least one original influencer tweet. Profile information for accounts which engaged with influencer content was also collected.
This dataset affords new opportunities for understanding radical right activity on Twitter, and is made publicly available for research use.
\footnote{The R\textsc{ot} dataset will be released shortly by the authors.}

\section{Related Work}
Radical right political activity represents a democratic risk for society. Far right actors routinely spread hateful, false messages which undermine press freedom, democratic institutions, and the rule of law. At worst, radical right activity can even facilitate terrorist attacks and other acts which cause physical harm \citep*{Koehler2019}.
The notoriously complex study of right wing political activity is beset by widespread ``terminological confusion'' and an ongoing ``war of words'' over definitions \citep*{mudde_2007, mudde1997}. Some have urged explicitly for the lack of ``an unequivocal definition" to be addressed by contributors to the field, but many definitional and conceptual challenges for those studying radical right activity remain unresolved \citep*{Carter2018}.
A range of competing terms are available to collectively describe right wing political activity, including ``far right'', ``extreme right'' and ``alt right''.
Notwithstanding the need for analytical clarity, terminological debate can distract scholars from empirical investigation. We adopt a broad understanding of the phenomenon captured by the term ``radical right'', as defined by Jupskås at the Centre for Research on Extremism: ``Right-wing radicalism can be defined as a specific ideology characterised by ``illiberal opposition to equality". It is associated with radical nationalism, authoritarianism, populism, and xenophobia.'' \citep*{Jupskas2020}.
Similarly, Mudde characterises the primary ideological aspects of the radical right as populism, nativism and authoritarianism \citep*{mudde_2007}.

In recent years, radical right activity has changed to reflect its adherents' pursuit of different goals. Online activity has increased, and new forms of communication and organisation have been adopted. 
Radical right influencers have increasingly become associated with the proliferation of misinformation about elections, public health, the environment, and the media \citep*{bennett2018}.
\cite{gattinara} argues that extreme right parties are best characterised as a social movement rather than as party political, due to a surge in ``extra-parliamentary grassroots activism''. The emergence of hybrid organisations—active online and in street politics, yet not contesting elections \citep*{Reid2020}—has accelerated this shift.
Organisation levels vary across the European radical right. In some European countries, highly organised parties of the radical right—such as the Freedom Party of Austria, the National Rally in France and the Dutch Party for Freedom—have contested elections with moderate success. By contrast, other countries such as the UK have almost no elected radical right politicians in the main parliamentary chamber.

The radical right has readily adopted digital technologies such as social media platforms and forums for the purpose of building communities \citep*{hartzell, Bright2020} attracting, radicalising, and communicating with supporters and members \citep*{sakki2016, scrivens2020}, and producing propaganda that can reach large audiences \citep*{klein2019, holt2020}.
Many radical right leaders, activists, parties, movements, and organisations have developed a strong presence on social media, exploiting its low cost and ease of access in order to bypass traditional media and engage directly with audiences \citep*{schroeder}. 
For example, 
the notoriety of Tommy Robinson (a former leader of the English Defence League) owed itself in part to his amateur production of inflammatory and prejudicial online content. 
Many radical right actors have also become increasingly active on niche and alternative platforms emphasising freedom of speech and minimal content-moderation, such as Gab, Parler and Voat \citep*{doi:10.1177/14614448211024546}.

Despite some countermeasures against harmful content—Twitter, Facebook, and Instagram have each banned Tommy Robinson from their platforms for instance—Youtube, Facebook, and several other large platforms continue to face the accusation that their users may follow dangerous``rabbit holes'' which promote and amplify extremist content \citep*{doi:10.1177/0894439314555329, 10.1145/3201064.3201081}.
Many commentators have argued that recommender systems prioritise radical right content simply because it attracts high rates of engagement (from both supporters and opponents), thereby creating ``radicalization pathways" \citep*{10.1145/3351095.3372879}. Content is promoted in order to optimise user experiences and maximise engagement, regardless of whether or not it creates risks to wellbeing \citep*{doi:10.1177/0163443718813467, bhargava_velasquez_2021}. Although prior research finds no evidence that Youtube's recommendation system leads to user radicalization, users who consume far right content have been found to exhibit higher engagement rates and greater ``stickiness" than users who consume any other type of content \citep*{Hosseinmardie}.
However, a precise understanding of the rationale each platform uses to prioritise and amplify its content can prove difficult to obtain, since the internal processes of social media companies are rarely left open for public scrutiny. 
 \citep*{Facebook2020}

Besides existing research into radical right content, several studies (e.g.,  \citet{Vosoughi1146}) have examined the characteristics of viral content generally, finding that misinformation can proliferate more readily than genuine information because of users' greater propensity to share novel or emotionally-charged content. 
However, accurate prediction of ``viral'' \textit{ex-ante} phenomena on social media remains challenging, especially in cases where information about the social system and content in question is incomplete \citep*{Martin2016}. 
Others have studied the factors driving engagement by examining the characteristics of users most likely to spread it.
\cite{Woolley2019} show how bots can increase the reach of political content, artificially widening its online audience by attracting inauthentic engagement.
\cite{Ozalp2020} investigate the online propagation of anti-Semitic content, finding that counterspeech against hateful narratives both spreads further and lasts longer than antagonistic content.
\cite{Becker10717} argue that social media platforms like Twitter tend to be centralised, where very few users hold the majority of the connections, and most users are disconnected from each other. These key influencers produce partisan content and amplify polarisation, and increasing their own influence.
To date, tandem consideration of extremist influencers and content remains underexplored, which is surprising given that both factors already independently help explain how content reaches large audiences.

\begin{landscape}
\begin{table}[htbp]
\scriptsize
\caption{Key information about the 35 radical right influencers. Mean engagement is the average number of retweets, quotes and replies per original entry. Original entries is the number of tweets, quoted tweets and replies produced by influencers. Number of  followers is recorded 9 January 2021. Media includes images, videos and gifs.}
\label{tab:influencer_overview}
\begin{tabular}{|p{4.3cm}|p{1cm}|p{1.3cm}|p{1.3cm}|p{1cm}|p{1cm}|p{1cm}|p{2cm}|p{1.1cm}| p{1.2cm}|p{1.5cm}|p{1.3cm}|}
 \toprule
    \textbf{Name} &  \textbf{Verified} & \textbf{Number of followers} & \textbf{Number of Original \newline entries} & \textbf{\% tweets} & \textbf{\% quotes} & \textbf{\% replies} & \textbf{Tweet mean toxicity (-/+1 SD)} &
    \textbf{\% tweets which contain media} & \textbf{\% tweets which contain URLs} &\textbf{Mean \newline engagement per original entry} &\textbf{\% followers who engage}\\
 \toprule
Altnewsmedia&FALSE&26,147&6,498&12.3\%&30.1\%&57.6\%&0.23 (0.03, 0.44)&3.2\%&10.0\%&18&24.9\%\\
Arktos Media&FALSE&10,553&80&82.5\%&1.3\%&16.3\%&0.1 (0.02, 0.18)&32.5\%&85.0\%&6&1.3\%\\
Arron Banks&FALSE&68,349&304&1.3\%&98.0\%&0.7\%&0.19 (0, 0.4)&0.7\%&4.9\%&211&12.3\%\\
Bill Warner&FALSE&36,605&203&64.5\%&0.5\%&35.0\%&0.32 (0.17, 0.47)&0.5\%&69.0\%&31&3.4\%\\
Breitbart London&FALSE&92,968&2,081&100.0\%&0.0\%&0.0\%&0.22 (0.06, 0.38)&0.1\%&104.9\%&51&10.6\%\\
British National Party (BNP)&FALSE&12,084&27&96.3\%&0.0\%&3.7\%&0.07 (0.03, 0.11)&0.0\%&103.7\%&9&0.4\%\\
Candace Owens&TRUE&2,716,743&468&39.3\%&33.1\%&27.6\%&0.3 (0.06, 0.54)&2.8\%&10.3\%&12,385&20.3\%\\
Catherine Blaiklock&FALSE&18,251&648&55.6\%&43.2\%&1.2\%&0.25 (0.06, 0.44)&3.9\%&37.3\%&165&30.5\%\\
Charlie Kirk&TRUE&1,839,724&1,720&97.2\%&2.6\%&0.3\%&0.23 (0.05, 0.42)&2.3\%&10.0\%&9,461&24.5\%\\
Daniel Friberg&FALSE&5,389&14&0.0\%&21.4\%&78.6\%&0.13 (0.04, 0.22)&28.6\%&50.0\%&18&1.4\%\\
David Vance&TRUE&172,014&3,416&39.0\%&49.9\%&11.1\%&0.18 (0, 0.35)&13.7\%&31.1\%&118&19.0\%\\
Edward Dutton&FALSE&11,178&766&42.6\%&9.9\%&47.5\%&0.18 (0, 0.35)&3.5\%&36.7\%&16&17.9\%\\
Elisabeth Hobson&FALSE&9,036&732&30.6\%&5.1\%&64.3\%&0.14 (0, 0.29)&14.8\%&36.3\%&4&5.0\%\\
Gerard Batten&FALSE&73,560&5,060&11.1\%&15.3\%&73.6\%&0.19 (0, 0.38)&5.1\%&12.8\%&104&27.2\%\\
Henrik Palmgren&FALSE&27,729&767&43.9\%&36.4\%&19.7\%&0.21 (0, 0.43)&12.6\%&53.7\%&30&12.9\%\\
James Delingpole&FALSE&68,856&4,132&10.5\%&25.1\%&64.4\%&0.22 (0, 0.46)&2.4\%&7.5\%&39&18.0\%\\
Lana Loktef&FALSE&54,221&954&19.3\%&44.0\%&36.7\%&0.28 (0.05, 0.52)&0.6\%&16.7\%&56&13.3\%\\
Laura Towler&FALSE&27,128&1,823&11.1\%&9.1\%&79.9\%&0.15 (0, 0.3)&3.5\%&8.3\%&36&21.0\%\\
Leave.Eu&TRUE&293,157&678&99.1\%&0.9\%&0.0\%&0.24 (0.05, 0.43)&0.9\%&5.5\%&806&10.9\%\\
Marcus Follin (The Golden One)&FALSE&21,084&226&27.9\%&34.5\%&37.6\%&0.14 (0, 0.29)&2.2\%&32.7\%&17&5.5\%\\
Mark Collett&FALSE&60,504&238&92.0\%&6.3\%&1.7\%&0.25 (0.02, 0.47)&15.5\%&66.4\%&457&19.9\%\\
Mark Meechan (Count Dankula)&TRUE&253,561&5,392&15.9\%&18.2\%&65.9\%&0.3 (0, 0.6)&15.3\%&20.3\%&99&22.5\%\\
Micheal Heaver&TRUE&64,319&527&76.9\%&14.0\%&9.1\%&0.17 (0.01, 0.33)&10.6\%&76.3\%&201&14.8\%\\
Morgoth&FALSE&13,317&424&9.9\%&13.4\%&76.7\%&0.22 (0, 0.45)&9.7\%&16.5\%&28&16.5\%\\
Nick Griffin&TRUE&38,324&973&88.8\%&2.7\%&8.5\%&0.24 (0.02, 0.46)&36.6\%&74.6\%&63&13.0\%\\
Nigel Farage&TRUE&1,668,203&515&75.5\%&24.3\%&0.2\%&0.16 (0.03, 0.29)&19.4\%&41.4\%&3,686&9.9\%\\
Pamela Geller&TRUE&199,301&3,949&74.9\%&24.0\%&1.0\%&0.28 (0.05, 0.51)&20.7\%&73.4\%&145&20.2\%\\
Paul Joseph Watson&TRUE&1,169,074&2,355&65.9\%&21.1\%&13.0\%&0.22 (0, 0.44)&13.3\%&56.1\%&956&18.4\%\\
Politicalite&FALSE&16,904&1,781&86.3\%&1.7\%&12.0\%&0.19 (0.03, 0.34)&13.0\%&85.6\%&50&28.0\%\\
Red Ice&FALSE&53,920&196&93.4\%&2.6\%&4.1\%&0.16 (0.04, 0.29)&15.8\%&94.4\%&28&2.6\%\\
Reform UK&TRUE&222,240&160&100.0\%&0.0\%&0.0\%&0.1 (0.05, 0.16)&18.8\%&76.9\%&247&4.0\%\\
Traditional Britain Group (Tbg)&FALSE&11,719&1,476&63.8\%&19.8\%&16.5\%&0.2 (0.02, 0.37)&27.0\%&57.6\%&22&20.8\%\\
Turning Point Uk&FALSE&35,895&924&50.3\%&41.2\%&8.4\%&0.18 (0.02, 0.34)&40.5\%&51.6\%&38&12.5\%\\
UK Independence Party&TRUE&206,236&193&94.3\%&4.7\%&1.0\%&0.17 (0.05, 0.29)&13.5\%&88.1\%&85&1.2\%\\
Westmonster&TRUE&76,249&629&99.5\%&0.0\%&0.5\%&0.13 (0.03, 0.23)&37.2\%&105.1\%&77&8.7\%\\
    \hline
    Mean&-&276,415&1,438&56.3\%&18.7\%&25.0\%&0.20 &12.6\%&48.9\%&850.3&14.1\%\\
std&-&608,839&1,692&0.3&0.2&0.3&0.06 &0.1&0.3&2,619&0.1\\
Total&-&9,674,542&50,329&41.8\%&21.4\%&36.8\%&-&10.6\%&35.4\%&-&-\\
    \bottomrule
\end{tabular}
\end{table}
\end{landscape}

\newpage
\section{Data}
\paragraph{Radical right influencers}
We identified 166 radical right influencers by manually reviewing the 2020 ``State of Hate'' report \citep*{HopeNotHate2020}. 84 had Twitter accounts.
We collected the profile information for all 84 accounts and retained only active public accounts with more than 5,000 followers, leaving 35.
We refer to these 35 accounts as ``influencers'', who are described in Table~\ref{tab:influencer_overview}.
During the period studied (see below) three influencers were suspended from Twitter: Lives Margoth on 27 July 2020, David Vance on 9 September 2020, and Mark Collett on 9 November 2020. 

\paragraph{Original entries from influencers}
From 8 July 2020 to 9 January 2021, we collected all content produced by the 35 influencers. 
For the multilevel analysis (see below), we exclude the retweets, as they were not originally created by the influencers and their overall engagement is likely to be primarily the result of other factors, such as the influence of the tweet's original creator. We also removed original entries from influencers during the final week of data collection. We choose a 1-week time window since original entries typically receive 99\% of their engagements within this time period, so by doing so we mitigate the risk of novelty bias in our analyses. In total, we collected 50,329 original entries, including posts (n = 21,037 posts, quote tweets\footnote{Quotes are retweets where the retweeter also makes a comment.} (n = 10,756), and replies (n = 18,536).


\paragraph{Engagement with influencers}
There are two main ways that users can engage with content on Twitter.
First, they can indicate their approval or support by selecting ``Like''.
Twitter's API provides information about the number of Likes each piece of content receives, but does not show which accounts Liked the content. This severely constrains how useful Likes can be for our analysis.
Second, users can share content by retweeting or quoting it, or by replying to it.
To understand how audiences engage with the influencers' content, we only consider the second form of engagement (Sharing). 
To check the robustness of this decision, we recollected all available original entries at the end of the data collection period and compared how many Likes and Shares they had received. 
The Pearson's correlation coefficient is 0.92, indicating a very strong linear association.
\footnote{Due to account suspensions, 17,697 original entries were no longer available at the end of the period. Otherwise, only 2.6\% of original entries had been deleted.}

We collected 29,760,855 engagements with the original entries, comprising retweets (n = 24,431,690), quotes (n = 383,574) and replies (n = 4,945,591).
\footnote{Note: we only collect content from public accounts on Twitter.}
Mentions of the influencers are not used for the multilevel analysis (see below), as they are not associated with original entries produced by the influencers.
These engagements were produced by 2,168,567 users, whom we refer to as ``engagers''.

\paragraph{Radical Right On Twitter Dataset (R\textsc{ot})}
All of the data used in this paper are made available as part of the Radical Right On Twitter dataset R\textsc{ot}.
R\textsc{ot} also contains other data which are not used here, including the lists of followers and friends of influencers, collected once per day for each influencer, as well as profiles and lists of followers for accounts which engaged with influencers' content.
See the short Data sheet in the Appendix.



\section{Methods}
To address our research question, we set the dependent variable to the count of engagements received by each entry, comprising the total number of retweets, quotes and replies.

\paragraph{Multilevel model}
Our data are ``nested'' in the sense that the 50,329 original entries are produced by influencers, and it is very likely that the number of engagements for original entries from the same influencer will be correlated, thereby violating the independence assumption of a single level model.
As such, we use a multilevel random-intercept model that allows us to represent unobserved influencer characteristics which affect engagement for each tweet. The simplest random-intercept model can be specified as follows: 
\[ y_{ij} = \beta_0 + \beta_1 x_{ij}+\mu_j+e_{ij} \]
where $i$ is the Level 1 (i.e. the tweet) and $j$ is the Level 2 (i.e., the influencer). $\beta_0$ is the model intercept, which varies by group according to $\mu_j$, the group effect. The random part of the model is comprised by $\mu_j + e_{ij}$, while the fixed part resides in $\beta_0 + \beta_1x_{ij}$. The inclusion of residuals at both Levels 1 and 2 allows us to estimate both tweet- and influencer-level variables, minimizing the risk of confounding.
Because the dependent variable is a highly skewed and over-dispersed count of engagements (see Table~\ref{tab:influencer_overview}) we expand the standard model to fit a negative binomial random-intercept model:
\[y_{ij} \sim NBD(\pi_i).\]
This model adds an explicit error term to account for over-dispersion, and therefore includes two parameters for the Level 1 variance, such as: 
\[y_{ij} \sim (\pi_i) + e_{0ij}z_{0ij} + e_{0ij}z_{1ij}\]
\[log(\pi_i) = \beta_0 + \beta_1x_{ij}+\mu_j\] where:
\[z_{0i} = \hat{\pi}_{i}^{0.5} ;  z_{1i} = \hat{\pi}_{i}\]
\[\mu_{0j} \sim N(0,\sigma^2_{u0})\]
where $\pi_i$ is the mean occurrence rate.
The model is implemented in R using the package glmmTMB \citep*{glmmTMB}.

We group-mean centered entry-level variables which are numeric, allowing us to isolate both within- and between-effects of variables on engagement.
Within-effects capture the effect of a higher or lower value of a variable relative to each influencer's usual activity, and between-effects capture the effect of a higher or lower value of a variable for each influencer on average. For instance, while the former can identify whether, given the typical toxicity of tweets produced by an influencer, producing a more toxic tweet leads to more engagement, the latter captures whether influencers who usually produce more toxic tweets receive more engagement. 
Variables are standardized to let us compare their coefficients.

\paragraph{Tweet-level independent variables}
Tweet-level variables are ``Level  1'' of the multilevel model (i.e., they correspond to original entries, which are the lowest unit of our analysis).
We construct a categorical variable for the type of content (i.e., whether it is original content, quote or reply).
Additionally, for each original entry we record typographical features, including the number of exclamation and question marks, number of hashtags, number of characters, and whether the entry contains URLs or rich media (e.g., photos, videos or gifs).
We also record semantic features by counting the number of positive and negative words using the NRC lexicon, and applying the toxicity classifier from Jigsaw's Perspective API\footnote{\url{https://www.perspectiveapi.com/}}.
The API returns a score from 0 to 1 for the toxicity of content.

Finally, we use Latent Dirichlet Allocation (LDA) topic modelling to infer the thematic content of the original entries. LDA is an unsupervised learning technique that treats the content of each original entry (or any `document') as a distribution over topics, and each topic as a distribution over words. The model is then optimized to find the parameters for the best combination of topics and words in each document \citep*{Blei2003}. We implement LDA in R using the textmineR package \citep*{textmineR}.
We selected 40 topics as the best fit, with alpha set to 0.1 and beta set to 0.01. See the Appendix for more information about LDA model fitting. 
37 of the 40 topics could be labelled and 9 were duplicates which we combined, leaving 28 unique topics.

\paragraph{Control variables}
Control variables are tweet-level independent variables which account for the circadian- and weekly-temporal rhythms of online activity. We include variables for the hour and day of the week when the original entries were produced.
They are also situated at ``Level 1'' of the multilevel model.

\paragraph{Influencer-level independent variables}
Influencer-level variables are ``Level  2'' of the multilevel model (i.e., they correspond to the influencers which create the original entries, reflecting the data's nested structure).
We split variables at the influencer-level into two groups: (1) followership size and (2) followership composition. 
\textbf{Followership size} comprises one variable, and measures the total number of followers that each influencer had on the day before they posted each original entry.
\textbf{Followership composition} measures the nature of the influencers' followers. We construct five variables: the mean number of statuses of influencers' followers; the mean number of followers of influencers' followers; the mean number of friends of influencers' followers; the percentage of influencers' followers who are suspended; deleted. 
Many of these measures are highly correlated (e.g., average number of friends and followers has a very strong association (p=0.91)). After testing different combinations of variables, we only include two in our models: the followers' mean number of statuses (i.e., how many tweets, retweets, quotes and replies they have produced) and percentage of their followers who are verified.

\section{Results}
\subsection{Exploratory Analysis}

\begin{figure}[!t]
    \centering
    \includegraphics[width = \columnwidth]{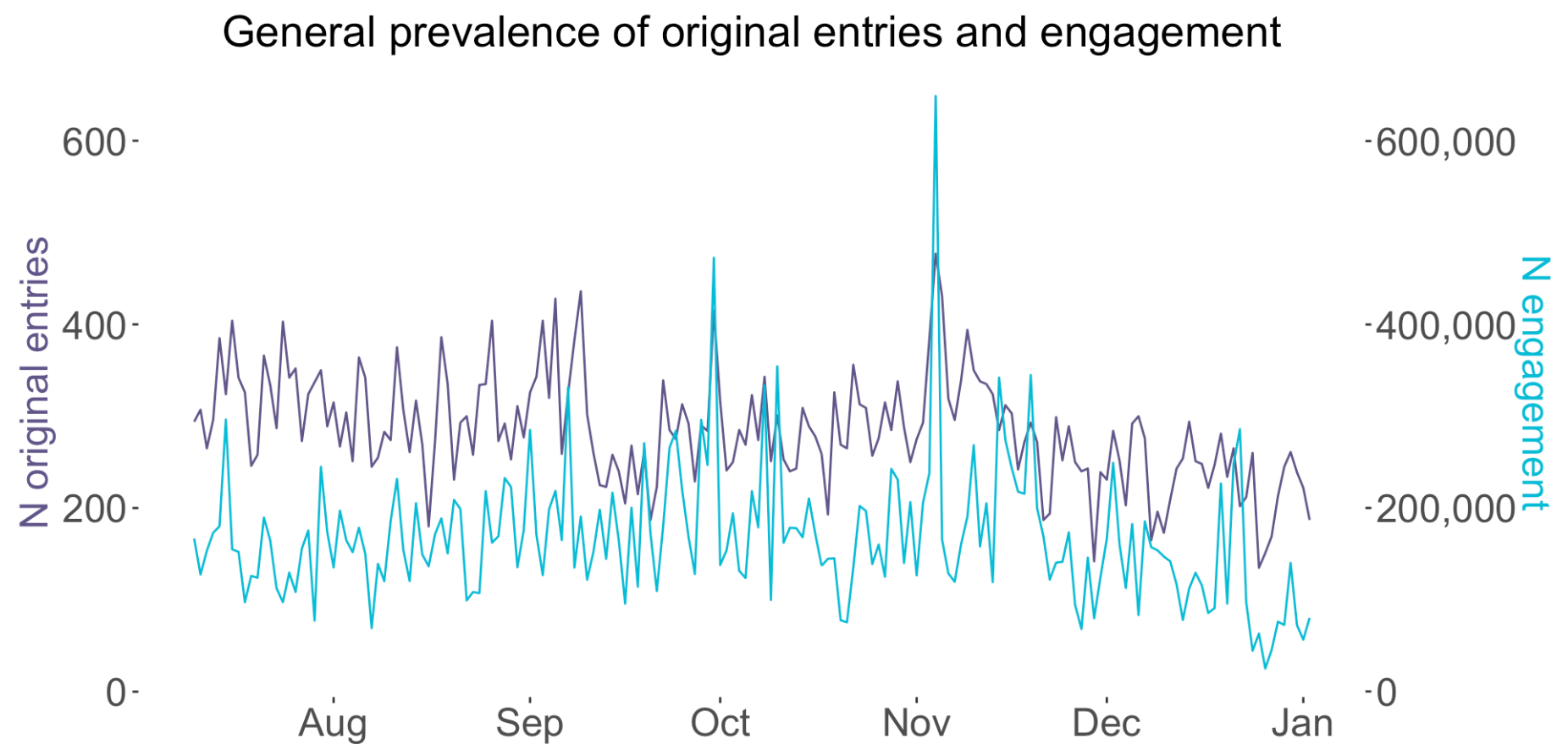}
    \caption{Original entries produced by influencers and total engagement.}
    \label{fig:time_series}
\end{figure}

\paragraph{Influencer activity}
Table~\ref{tab:influencer_overview} shows key statistics about influencers' activity.
There is a high degree of variation, with influencers producing between 14 original entries (Daniel Friberg) and 6,498 original entries (Altnewsmedia).
Influencers also differ in how they use Twitter. Some influencers, such as Aaron Banks, primarily produce quotes (accounting for 98\% of his activity), with very few original tweets. Others, such as Breitbart London, only produce posted tweets (accounting for 100\% of theirs).
Influencers' numbers of followers vary widely. 
Candace Owens has 2.67 million followers, and Charlie Kirk and Nigel Farage have over 1.6 million. In contrast, Traditional Britain Group has only 11,719, and Elisabeth Hobson has only 9,036. This reflects the very different sizes of potential audience to which each influencer has access, and likely reflects very different levels of engagement.
The 35 influencers produced content consistently during the period, with two notable spikes.
First, during the first US presidential debate on 29 September 2020, and second, during the US presidential election on 3 November 2020.
The number of original entries and engagements over time is shown in Figure~\ref{fig:time_series}.

\paragraph{Influencer engagement}
On average, across all influencers, 95\% of engagement occurs within 6.86 hours, and 99\% within 9.19 hours. 99.99\% of engagements was received within one week. 
On average, Candace Owens received the most engagements, with 12,385 engagements for each original entry, and Elisabeth Hobson received the fewest engagements, with 4 engagements on average.
10.5\% of the original entries received no engagement. As expected, the average number of engagements for each influencer is highly correlated with their number of followers (p=0.93).
Interestingly, influencers vary substantially in how many of their followers engage with their content. For some influencers, only  1\% of their followers engage with them (e.g., Arktos media and UK Independence Party), whereas others have 28-30\% (e.g., Catherine Blaiklock and Politicalite). This indicates very different levels of interest from their followers.
The number of followers and \% of followers who engage is positively correlated, but only weakly (p = 0.18).

\paragraph{Thematic analysis of content}
\begin{figure}[htbp]
    \centering
    \includegraphics[width = \columnwidth]{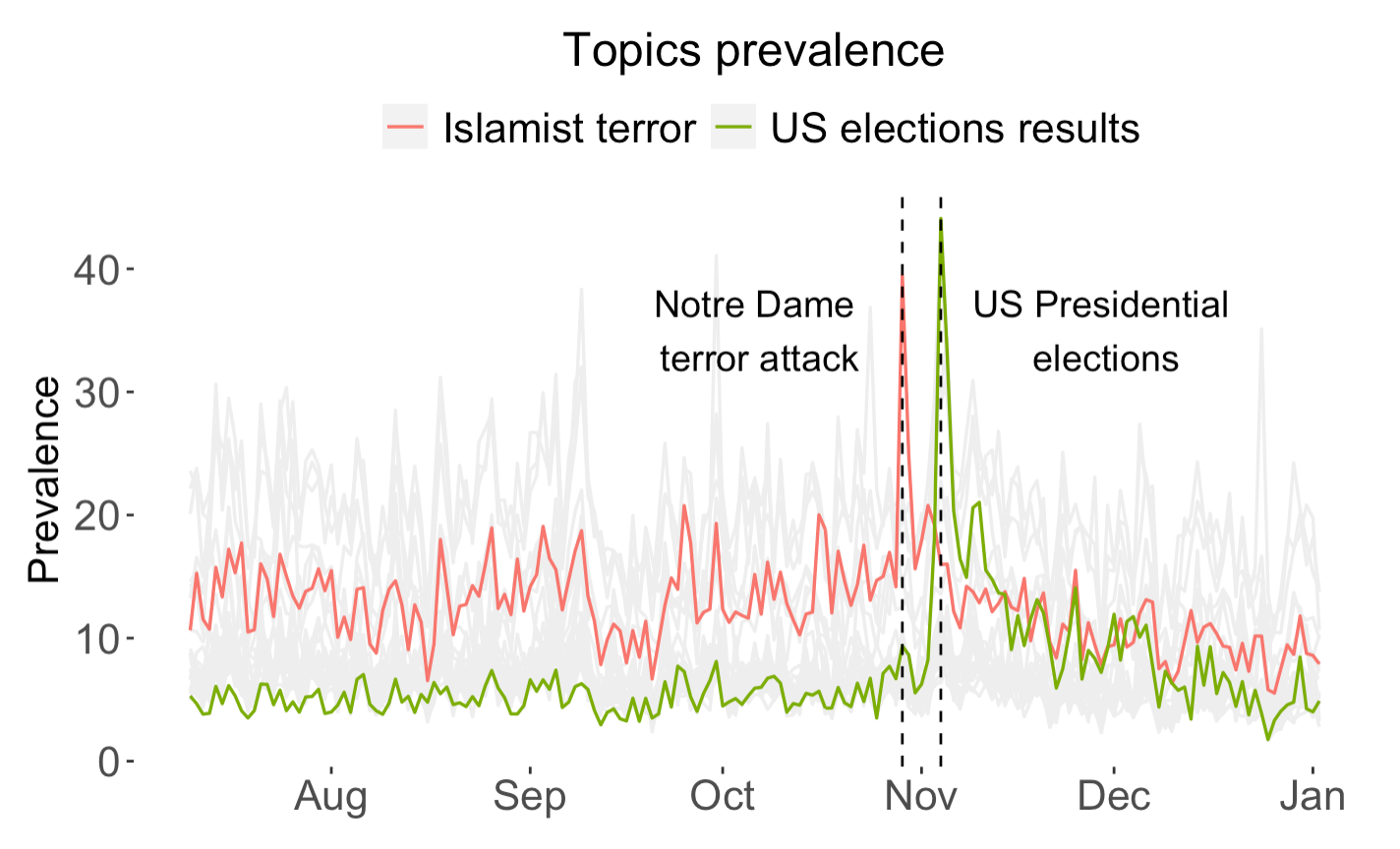}
    \caption{Topics prevalence.}
    \label{fig:topics_prevlence1}
\end{figure}
The content produced by the influencers has an average toxicity of 0.2. We noticed that some influencers directly request retweets to drive engagement (e.g. "RT if you agree!"), which occurred in 218 tweets and 7 quoted tweets (0.45\% of the total).
While 35.4\% of original entries contain a URL, a small number of entries contain rich media, such as images (8.81\%), gifs (0.49\%) and videos (1.53\%).
The 28 topics show temporal variations which can in some cases be linked to external events. In Figure~\ref{fig:topics_prevlence1} we highlight two topics, `Islamist terror' and `US elections results'. These peaked following the Notre Dame terror attack at the end of October 2020, and the US Presidential elections in November, respectively.

\begin{table*}[htbp]
\caption{Random effect models comparison.}
\begin{tabular}{lllccc}
 \hline
     &Model&\multicolumn{1}{|p{2cm}|}{\centering AIC \\ (Difference with AIC)}&\multicolumn{1}{|p{2cm}|}{\centering Level 1 \\ Unexplained variance}&\multicolumn{1}{|p{2cm}|}{\centering Level 2 \\ Unexplained variance}&Adjusted ICC\\
    \hline
    \hline
1&No predictors&507,345&1.24&3.63&0.74\\
2&content (minus topics)&476,604 (-30,741)&0.90&1.73&0.66\\
3&topics&505,066 (-2,279)&1.22&3.42&0.74\\
4&all content&475,049 (-32,296)&0.89&1.69&0.66\\
5&followership composition&507,113 (-232)&1.24&3.37&0.73\\
6&followership size&507,092 (-253)&1.24&1.38&0.53\\
7&composition + size&507,096 (-249)&1.24&1.40&0.53\\
8&all content + all follower vars&475,003 (-32,342)&0.89&0.44&0.33\\
    \hline
\end{tabular}
\end{table*}

\subsection{Model Results: Understanding Engagement With Original Entries}

Table~\ref{tab:model_results} shows goodness of fit results and variances for a baseline random intercept model with no independent variables, as well as seven models that include different combinations of content-level (``Level 1'') and influencer-level (``Level 2'') variables.
To compare overall model fit we use Akaike Information Criterion (AIC), a measure that penalises more complex models and rewards more parsimonious models.
We also report the Level 1 and Level 2 variances.
Level 1 variance is the unexplained variation at Level 1 (i.e., the original entries), after controlling for the independent variables specified in the model. It can be interpreted as a Level 1 error term. A lower value indicates that the model has explained more of the tweet-level features which drive engagement.
Level 2 variance is the unexplained variation at the grouping level (i.e., the influencer) after accounting for the independent variables. It can be interpreted as a Level 2 error term, where lower Level 2 variance indicates that the model has explained more of the influencer-level features which drive engagement. 
When additional explanatory variables are added to a model it is expected that Level 1 variance will decrease. However, Level 2 variance might increase or decrease depending on whether the distribution of the variable differs across the Level 2 grouping unit (i.e., influencers).
We also look at the Adjusted Intra-class coefficient (ICC).
The Adjusted ICC measures the percentage of unexplained total variance (Level 1 + Level 2 unexplained variance) which can be attributed to the grouping factor alone. It is calculated as Level 2 variance / (Level 1 variance + Level 2 variance).

Model 1 is our baseline model, with no independent variables. The Adjusted ICC is 0.74, which means that the influencer variable accounts for 74\% of total variance, confirming the highly nested structure of our data and the high variation in how much engagement influencers receive. 
When we add content variables (minus topics) to Model 1 (Model 2), model fit improves, with AIC reduced by 30,741. Interestingly, both Level 1 and Level 2 unexplained variation decrease considerably (1.24 to 0.9 and 3.63 to 1.73, respectively), which suggests that influencers differ in the type of content they produce.
Model 3 adds topic variables to Model 1, obtaining a far more modest improvement compared with Model 2, with a much smaller reduction in AIC (only 2,279).
Model 4 includes all content variables, and shows the greatest improvement in fit, with the AIC reducing by 32,296. 

Model 5 introduces the followership composition variables to Model 1, which only vary at influencer level. The AIC only marginally reduces compared with Model 1 (by 232). The inclusion of this Level 2 variable has no effect on unexplained Level 1 variance (which stays stable at 1.24) and little effect on level 2 variance, reducing it from 3.63 to 3.37. 
Model 6 includes the number of followers to Model 1. Even though the AIC also reduces only marginally compared with Model 1 (by 254), the unexplained random variance is substantially reduced, falling to 1.38. This shows the role of followership size in explaining how many engagements influencers' tweets receive.
Nevertheless, when including all followership variables (Model 7) overall fit gets worse compared to Model 6, even though it does show improvements with respect to the baseline model.
Finally, Model 8 includes all of the content and followership variables, with the unexplained Level 1 variance reduced by 29\% (1.24 to 0.89) and the unexplained Level 2 variance by 88\% (3.63 to 0.44).

Table~\ref{tab:model_results} shows the coefficients for Model 8.
Coefficients are expressed as Incidence Rate Ratios (IRR), a measure of the change in the dependent variable (the engagement) with every unit increase in the independent variable. Values above 1 indicate an increase in the incidence ratio, and below 1 a decrease.
We observe that most content variables are significant.
The type of variable has a large impact, where compared with tweets, quoted entries are associated with less engagement (IRR = 0.67), as are replies (IRR=0.05).
Inclusion of media also makes a difference to engagement. Original entries with a URL or GIF receive less engagement (IRR=0.79 and IRR=0.53). In contrast, entries with a video or image receive more engagement (IRR=3.20 and IRR=1.67).
Also, tweets that request retweets receive substantially more engagements (IRR = 2.61).
Interestingly, there are positive within- and between-effects for toxicity, suggesting not only that more toxic tweets receive more engagement, but also that influencers who produce more toxic content on average receive more engagement. Moreover, the between-effect is greater than the within-effect (1.53 and 1.13, respectively).
Similarly, we show that there are positive within- and between-effects for the length of content, as measured by the number of characters. Within-effects have an IRR of 1.21 and between-effects have an IRR of 1.31. 
22 of the 29 topics are significant (p$<$0.05), although in most cases the IRRs are small (between 0.91 and 1.11). For brevity, we only show the range of IRRs associated with different topics in Table~\ref{tab:model_results}.
Numerous typographical factors, such as number of hashtags and exclamation marks, have significant but small effects. 
Finally, the followership composition variables (followers' mean statuses count and \% of followers which are verified) are both non-significant. This is surprising, and may reflect the size of our sample and the time period studied.
The number of influencers' followers has an IRR of 2.20, showing its importance in driving engagement.

\begin{table}[htbp]
\caption{Model 8 results. Note that 22 of the 29 topics are significant (p$<$0.05).}
\label{tab:model_results}
\begin{tabular}{llc}
 \toprule
    \textbf{Variable} & \textbf{IRR} &
    \textbf{SE}\\
 \toprule
\textbf{Content variables} & \\
\hline
Is a quote&0.67 ***&0.02\\
Is a reply&0.05 ***&0.02\\
Contains a url&0.79 ***&0.02\\
Contains an image&1.67 ***&0.02\\
Contains a video&3.20 ***&0.05\\
Contains gif&0.53 ***&0.09\\
Toxicity within&1.13 ***&0.01\\
Toxicity between&1.53 ***&0.12\\
Length within&1.21 ***&0.01\\
Length between&1.31 **&0.10\\
N exclamation marks within&0.99 *&0.01\\
N exclamation marks between&1.04&0.10\\
N question marks within&1.03 ***&0.01\\
N question marks between&1.14&0.11\\
N hashtags within&1.02 ***&0.01\\
N hashtags between&0.85&0.10\\
N positive words within&0.99 *&0.01\\
N positive words between&1.14&0.09\\
N negative words within&0.96 ***&0.01\\
N negative words between&0.79&0.15\\
Requests a Retweet &2.61 ***&0.09\\
Topic (all 29 topics)&0.91-1.11&0.01\\
\hline
\textbf{Followership variables} & \\
\hline
Followers' mean statuses count&0.77&0.18\\
\% verified followers&1.19&0.13\\
N followers&2.20 ***&0.11\\

\hline
\hline
\textbf{Goodness of fit} & \\
\hline
    -2 Loglikelihood & 481,112 & -\\
    Adjusted ICC &0.33 & -\\
    Conditional r2\footnote{As defined by \cite{Nakagawa}} & 0.84&- \\
    \bottomrule
    \footnotesize{Signif. codes:  0 ‘***’ 0.001 ‘**’ 0.01 ‘*'}
\end{tabular}
\end{table}

\section{Discussion}
\paragraph{What factors are associated with radical right content that receives a lot of engagement on Twitter?}
Our results demonstrate the importance of taking into account influencer-level factors, as shown by the Adjusted ICC of 0.74 in Model 1 (which only contains the influencer-level variable as part of the model structure).
The evidence on which influencer-level variables are the most important is mixed. Followership size is the most important influencer-level factor in Model 8, with a substantial IRR, and drives the largest reduction in Level 2 variance. Nevertheless, both of the variables associated with followership composition are not significant in Model 8, and do not show a significant impact as shown by comparing Model 5 with Model 6. Notably, the AIC for Model 7, which includes both followership size and composition, is worse than Model 6 which only includes followership size. This is likely because there is a relationship between followership size and composition, which reduces/eradicates the marginal benefits of including additional variables. Overall, these analyses indicate that followership size is important and that followership composition might matter, but the relationship is partly confounded by associations between followership size and composition. 
Inclusion of content-level variables leads to substantial increases in model fit. We demonstrate that the type of content (original posts, quotes and replies), the toxicity of content, and whether influencers request retweets are particularly important, followed by whether media is included, and a range of typographical and topic factors.

Model 8 shows that including both influencer- and content-level variables leads to the greatest improvement in model fit, with the lowest AIC and lowest Level 1 and Level 2 unexplained variances. This strongly indicates that both sets of variables are needed to explain engagement with radical right influencers' content.
It is worth noting that Model 8 still has unexplained variance at both Level 1 and Level 2, with the smallest reduction from Model 1 in the Level 1 unexplained variance. More advanced modelling, with more variables, would help to increase explanative power.
One limitation of our research design is that we do not investigate radical right content which is produced by unknown (or less known) figures yet still goes viral. It is plausible that for such content the dynamics of engagement are different. This means that we cannot generalise our results to \textit{all} radical right content, only to content produced by radical right influencers. In the future, this could be addressed by supplementing our influencer-based research design with a hashtag- or keyword-based design.

\paragraph{The relationship between engagement and followers}
Interestingly, we identify a substantial range in the \% of each influencers' followers who engage with that influencer's content. This should be explored in future work as it indicates the importance of what could be termed the `activation rate', and of whether influencers, especially those with large followings, can motivate their followers to engage with their content. This raises an interesting empirical question regarding whether followership drives engagement or engagement drives followership. This in turn invites a conceptual question about whether influencers aim to have a large audience (i.e., many followers) who might be exposed to (or subtly influenced by) the influencers' content, or aim instead for engagements, which can be seen as a better metric of genuine interest. We have started from the assumption that engagement rate is more important than audience size, but this decision could be explored further in future work.
Understanding the relationship between engagement and followers is particularly difficult because of the widespread presence of bots in online settings, since these might be used by influencers to artificially and inconsistently increase engagement, followership, or both.

\paragraph{Methodological limitations and future work}
There are several methodological limitations in the present study.
The dataset that we use covers only six months of data and 35 influencers. Our method could be extended over a longer period to enable more in-depth longitudinal analyses. The set of 35 influencers could also be extended, potentially enabling greater global coverage. There were a small number of data outages (reported in the Appendix). These are unlikely to have impacted our results, but could inhibit other analyses with R\textsc{ot}.
To assess the content of tweets, we use established computational tools including topic models, keyword dictionaries, and the toxicity classifier from Jigsaw's Perspective API.
Qualitative manual analysis of the tweets, either using a directed or undirected annotation framework, could provide more nuanced assessments of content. More advanced and tailored AI-based tools could also provide better measurement.
We only analyse Twitter content. Niche free-speech sites, such as Gab, Parler, and Voat are increasingly used by radical right influencers, and could therefore be an important complement to the mainstream platform content analysed here. This would be particularly useful for analysing the behaviour of radical right influencers who are banned from mainstream platforms like Twitter.
Finally, through our use of multilevel modelling, we have sought to explain the factors which drive engagement with radical right influencers' content. 
In future work, we could adapt the methods used here to predict aggregated user engagement, which would aid policymakers, regulators and platforms in understanding the generative process of hazardous viral content.

\section{Conclusion}
Few studies have adopted an influencer-based approach to analyse incendiary, divisive, and extremist online messages. We use this approach to show the importance of influencers in the radical right ecosystem. In addressing our research question, we have produced new insights into the activity of the radical right on Twitter and the factors which drive viral engagement with their content. 
We have also introduced R\textsc{ot}, a new resource for understanding radical right behaviour, which is made publicly available. 
Our results demonstrate the need for more innovative research designs, involving methods which directly account for the dynamics and structure of radical right behaviour online.

\bibliography{references.bib}
\bibliographystyle{aaai}


\section{Appendix: Topic models}
The original entries were pre-processed by lowering the case, removing stop words, punctuation, numerical characters, urls and \@ mentions, and stemming. Emojis and currency signs were not removed.
We fit the topic model sequentially, first optimizing the number of topics (k) with alpha and beta set to 0.1, and then fitting alpha and beta. Models fit was assessed with log-likelihood and average coherence. Model fit was cross-validated, splitting the data into five random groups, training the model on four and then evaluating on the fifth.
We manually reviewed models in the range of 15 to 50 topics by examining the most probable words and original entries with a high probability for the topic ($>$0.65). A model with 40 topics was selected, of which 37 could be labelled and 9 were duplicates, leaving 28 unique topics.

We follow the methodology described by \citet{Vidgen2020b} to validate the topic model. For overall coherence evaluation we use the "word intrusion" method, in which external subjects are presented with six words by topic (the five most probable words plus one of the least probable ones). We measure the percentage of times that the intrusive words were correctly identified. 
To test for overall validity of the topics, we present the subjects with fifty high probability tweets within each topic to review whether they are correctly assigned.
We then ask three students that are not familiar with this study to conduct both tests reaching an overall accuracy of 84\% for coherence, and 86\% for accuracy.

\section{Appendix: Datasheet}
\subsection{Motivation}
We collected the Radical Right On Twitter dataset (R\textsc{ot}) to advance research into radical right activity online.
The resource addresses a lack of data in this field, particularly data which relates to the activity of radical right influencers.
The dataset was funded without commercial support. 

\subsection{Composition}
R\textsc{ot} follows six months of Twitter activity (8th July 2020 to 9th January 2021) from 35 radical right influencers.
We follow the advice given by \cite{Williams2017} for publishing Twitter data on sensitive topics.
R\textsc{ot} includes:
\begin{itemize}
    \item \textbf{Influencers' content}: all content produced by the influencers, including posts (n = 22,131), replies (n = 19,947), quotes (n = 11,314), and retweets (n = 37,283).
    \item \textbf{Influencers' profiles}: Twitter profile information for all 35 influencers.
    \item \textbf{Influencers' followers}: a list of each influencers' followers, collected each day (combined n = 6,592,056). 
    \item \textbf{Influencers' friends}: a list of each influencers' friends, collected each day (combined n = 262,856).
    \item \textbf{Engagement}: all tweets which engage with influencers, including replies, quotes, retweets and mentions (n = 43,647,888). 
    \item \textbf{Engagers' followers}: List of followers of every user who replied, quoted or retweeted influencer content. We only collected users' list of followers once, even if they engaged with the influencers multiple times during the period studied.
\end{itemize}

\paragraph{Data outages}
When building R\textsc{ot}, we did not identify any data outages in the collection of original entries, engagement, mentions, and profiles.
On 14 days \footnote{July 10, 12, 13, 15, 17, 18, 20, 23, 25, 27, 29, 31 and August 5, 6} during the first month of collection the daily list of followers was not collected for 8 influencers\footnote{blaiklockBP, PrisonPlanet, DVATW, GerardBattenUK, LeaveEUOfficial, Michael\_Heaver, Henrik\_Palmgren, LanaLokteff}.
We verified that the between days variation was never above (0.2\%), and used the day prior to the data outage as a proxy.

\subsection{Collection process}
\paragraph{Influencer collection}
We identified 166 radical right influencers by manually reviewing the 2020 ``State of Hate'' report \cite{HopeNotHate2020}. 84 had Twitter accounts. We collected the profile information for all 84 accounts and retained only those which had active public accounts with more than 5,000 followers, leaving 35 accounts.
We refer to these 35 accounts as ``influencers''. 

\paragraph{Collection mechanism}
All data was collected in an AWS server using Twitter API v1 stream, profiles, friends and followers endpoints. 
During the period studied three influencers were suspended from Twitter: Lives Margoth on 27th July 2020, David Vance on 9th September 2020, and Mark Collett on 9th November 2020.

\paragraph{Ethics review}
The Alan Turing Institute's Ethical Advisory Group reviewed and approved the research.







\end{document}